\documentclass[%
 reprint,
 prc,
 amsmath,amssymb,
 aps,
 lengthcheck,
floatfix,
]{revtex4}

\usepackage{balance}
\usepackage{graphicx}
\usepackage{dcolumn}
\usepackage{bm}
\usepackage{hyperref}
\usepackage[utf8]{inputenc}

\usepackage{multirow}
\usepackage{soul}
\usepackage{float}
\usepackage{graphicx}
\usepackage{times}
\usepackage[normalem]{ulem}
\usepackage{color}
\usepackage{cellspace}
\usepackage{lipsum}

\newcommand{\be}{\begin{equation}}
\newcommand{\ee}{\end{equation}}
\newcommand{\bea}{\begin{eqnarray}}
\newcommand{\eea}{\end{eqnarray}}

\newcommand {\nonu}{\nonumber}

\newcommand{\comment}[1]{}
\renewcommand\sout{\bgroup \color{red} \ULdepth=-.5ex \ULset}
\def\simge{\mathrel{\rlap{\raise 0.511ex
     \hbox{$>$}}{\lower 0.511ex \hbox{$\sim$}}}}
\def\simle{\mathrel{\rlap{\raise 0.511ex
      \hbox{$<$}}{\lower 0.511ex \hbox{$\sim$}}}}

\begin{document}


\title{Direct mapping of tidal deformability to the iso-scalar and iso-vector nuclear matter parameters}

\author{\href{https://orcid.org/0000-0003-3308-2615}Sk Md Adil Imam$^{1,2}$\includegraphics[scale=0.06]{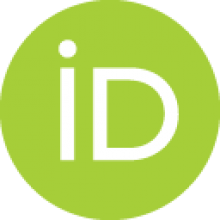}}
\email{mdadil.imam@saha.ac.in}
\author{\href{https://orcid.org/0000-0003-1274-5846}Arunava Mukherjee$^{1,2}$\includegraphics[scale=0.06]{Orcid-ID.png}}
\email{arunava.mukherjee@saha.ac.in}

\author{\href{https://orcid.org/0000-0001-5032-9435}B. K. Agrawal$^{1,2}$\includegraphics[scale=0.06]{Orcid-ID.png}
}
\email{bijay.agrawal@saha.ac.in}
\author{\href{https://orcid.org/0009-0002-7395-3198}Gourab \surname{Banerjee}$^{1,2}$\includegraphics[scale=0.06]{Orcid-ID.png}}

\affiliation{$^1$Saha Institute of Nuclear Physics, 1/AF 
Bidhannagar, Kolkata 700064, India.}  
\affiliation{$^2$Homi Bhabha National Institute, Anushakti Nagar, Mumbai 400094, India.}

\date{\today}

\begin{abstract} 
\begin{description}
\item[Background :]  The equations of state (EoSs) which determine the properties of neutron stars (NSs) are often characterized by the iso-scalar and iso-vector nuclear matter parameters (NMPs). Recent attempts to relate the radius and tidal deformability of a NS to the individual NMPs have been inconclusive. These properties display strong correlations with the pressure of NS matter which depends on several NMPs. It may be necessary to map the NS properties to the NMPs.

\item[Purpose :] To identify the important NMPs required to describe the tidal deformability of neutron star for astrophysically relevant range of their gravitational masses (1.2 -- 1.8 M$_\odot$) as encountered in the binary neutron star merger events.

\item[Method :] We construct a large set of EoSs using four iso-scalar and five iso-vector NMPs. These EoSs are employed to perform a systematic analysis to isolate the NMPs that predominantly determine the tidal deformability parameter obtained by solving the Tolman–Oppenheimer–Volkoff (TOV)  equations. The tidal deformability for the EoSs consistent with the chiral effective field theory ($\chi$EFT) at the lower density and satisfying maximum gravitational mass of stable NS $\geq$ 2 M$_\odot$ are directly mapped to these NMPs.

\item[Results :] We provide empirical relations between tidal deformability parameter and a minimal set of essential NMPs through lower order polynomial functions. The tidal deformability of the NS with mass 1.2-1.8 M$_\odot$ can be determined within 10-30\% directly in terms of linear function of four nuclear matter parameters, namely, the incompressibility coefficient $K_0$ and skewness $Q_0$ of symmetric nuclear matter, and the slope $L_0$ and curvature parameter $K_{\rm sym0}$ of symmetry energy. 
The inclusion of remaining NMPs improves the predictability of tidal deformability to 5-10\%.

\item[Conclusion :] Empirical relations are developed for quick estimation of reliable values of tidal deformability, in terms of the NMPs,  across a wide range of NS mass for a given EoS model . Our method can also be extended to other NS observables.
\end{description}
\end{abstract}

\maketitle

\section{Introduction}

The stringent constraints on the equation of state (EoS) of the dense matter promised by gravitational wave astronomy through the detailed analysis of Bayesian parameter estimation has triggered many theoretical investigations of the neutron star (NS) properties~\cite{GW170817, GW170817_EOS_prl2018, Malik2018, De18, Fattoyev2018a, Landry2019, Piekarewicz2019, Malik:2019whk, Biswas2020, Abbott2020, Dietrich_etal_Science2020, Thi2021}. The tidal deformability parameter ($\Lambda$) of NS,  which encodes the information for the EoS has been inferred from a gravitational wave event GW170817 observed with the Advanced-LIGO~\cite{LIGOScientific:2014pky} and Advanced-Virgo detectors~\cite{VIRGO:2014yos} from a binary neutron star (BNS) merger with component masses in the range 1.17--1.6 M$_\odot$~\cite{Abbott18a, Abbott2019}. Subsequently, another event GW190425, likely originating from the coalescence of BNSs was observed~\cite{Abbott2020}. The GW-signals from coalescing BNS events are likely to be observed more frequently in the upcoming runs of LIGO-Virgo-KAGRA and the future detectors, e.g., Einstein Telescope~\cite{Punturo:2010zz} and Cosmic Explorer~\cite{Reitze:2019iox}. Complementary information on the NS properties is also provided by the Neutron star Interior Composition Explorer (NICER). It relies on pulse profile modeling, a powerful technique to monitor electromagnetic emission from the hot spots located on the surface of the neutron star~\cite{Watts2016, Psaltis_2014}. Recently, two different groups of NICER have reported neutron star's mass and radius simultaneously for PSR J0030+0451 with radius R $=13.02^{+1.24}_{-1.06}$km for mass M $=1.44^{+0.15}_{-0.14}$ M$_\odot$~\cite{Miller2019} and R $=12.71^{+1.14}_{-1.19}$ km for M $=1.34^{+0.15}_{-0.16}$ M$_\odot$~\cite{Riley2019}, and for another (heavier) pulsar PSR J0740+6620, R $=13.7^{+2.6}_{-1.5}$ km with M $=2.08 \pm 0.07$ M$_\odot$~\cite{Miller:2021qha} and R $=12.39^{+1.30}_{-0.98}$ km with M $=2.072^{+0.067}_{-0.066}$ M$_\odot$ \cite{Riley:2021pdl}.  

The NS matter up to 2-3 times the saturation density ($\rho_0$ = 0.16 fm$^{-3}$) is expected predominantly to be composed of nucleons in $\beta$-equilibrium. The EoSs for such matter can be expressed using iso-scalar and iso-vector nuclear matter parameters which characterize the symmetric nuclear matter (SNM) and density-dependent symmetry energy, respectively. Several investigations have been carried out to narrow down the bounds on these NMPs from the information on the radius and tidal deformability of a canonical neutron star~\cite{Alam:2016cli,Carson:2018xri, Malik2018, Forbes_etal_2019prd, Tsang:2019vxn, Guven:2020dok, Malik:2020vwo, Tsang:2020lmb, Malik_book, Reed:2021nqk, Ghosh:2021bvw, Beznogov:2022rri,PRADHAN2023vm}. The behavior of EoSs around $\rho_0$ may be important in determining the properties of such NSs. Several studies have been performed to study the correlation between NS properties and nuclear matter parameters~\cite{Fortin16,Tsang:2020lmb,Thi2021}. However, the correlation between neutron star properties and individual nuclear matter parameters is found to be at variance~\cite{patra23}. But the correlations of NS radii with the pressure at the densities $\sim 2\rho_0$ for the $\beta$-equilibrated matter is found to be robust \cite{Lattimer:2000nx}. Similar trends are observed for the correlations of the tidal deformability with pressure at $\approx$ $2\rho_0$~\cite{Tsang:2019vxn, Tsang:2020lmb, Patra:2022yqc}. These correlations are found to be nearly model-independent and persist over a wide range of NS mass 1.2-2 M$_\odot$~\cite{Tsang:2020lmb,Patra:2022yqc}. This information has been  widely used to obtain empirical relation between the NS properties and pressure for $\beta$-equilibrated matter at density $\approx$ $1.5-2\rho_0$~\cite{Lattimer:2000nx,Lim_PRL,particles6010003, lim2023symmetry}. Nevertheless, these relations cannot constrain the underlying iso-scalar and iso-vector NMPs components separately~\cite{Imam:2021dbe,Chiranjib22,Tovar21}. It is therefore important to map the NS properties directly in terms of the NMPs, which describe the EoS.

In the present work, we perform systematic analysis and multi-parameter correlation studies to identify the minimal set of essential iso-scalar and iso-vector nuclear matter parameters that predominantly determine tidal deformability of neutron stars obtained from solutions of TOV equations with masses 1.2 -- 1.8 M$_\odot$\cite{Hinderer_2008}. Here we show that the tidal deformability can be mapped, predominantly to the two iso-scalar parameters incompressibility (K$_0$), skewness (Q$_0$) and the two iso-vector parameters slope (L$_0$), curvature (K$_{\rm sym0}$). The inclusion of a few higher order parameters improves the predictability of the mapped functions, in particular, for the heavier neutron stars.
Estimating the NMPs using Bayesian inference from the observational information of NS properties required to construct large number of EoSs from NS matter and corresponding numerical solution of the TOV equations, and to test their astrophysical validity (see second paragraph of section II) which is often computationally expensive~\cite{Bilby_ref,Riley,Brandes2023}. However, our established empirical relation can be directly employed to facilitate the Bayesian analysis of diverse astrophysical observations.

\section{Mapping Tidal deformability to the nuclear matter parameters}
\label{Results}

\begin{figure}
	\includegraphics[width=0.5\textwidth]{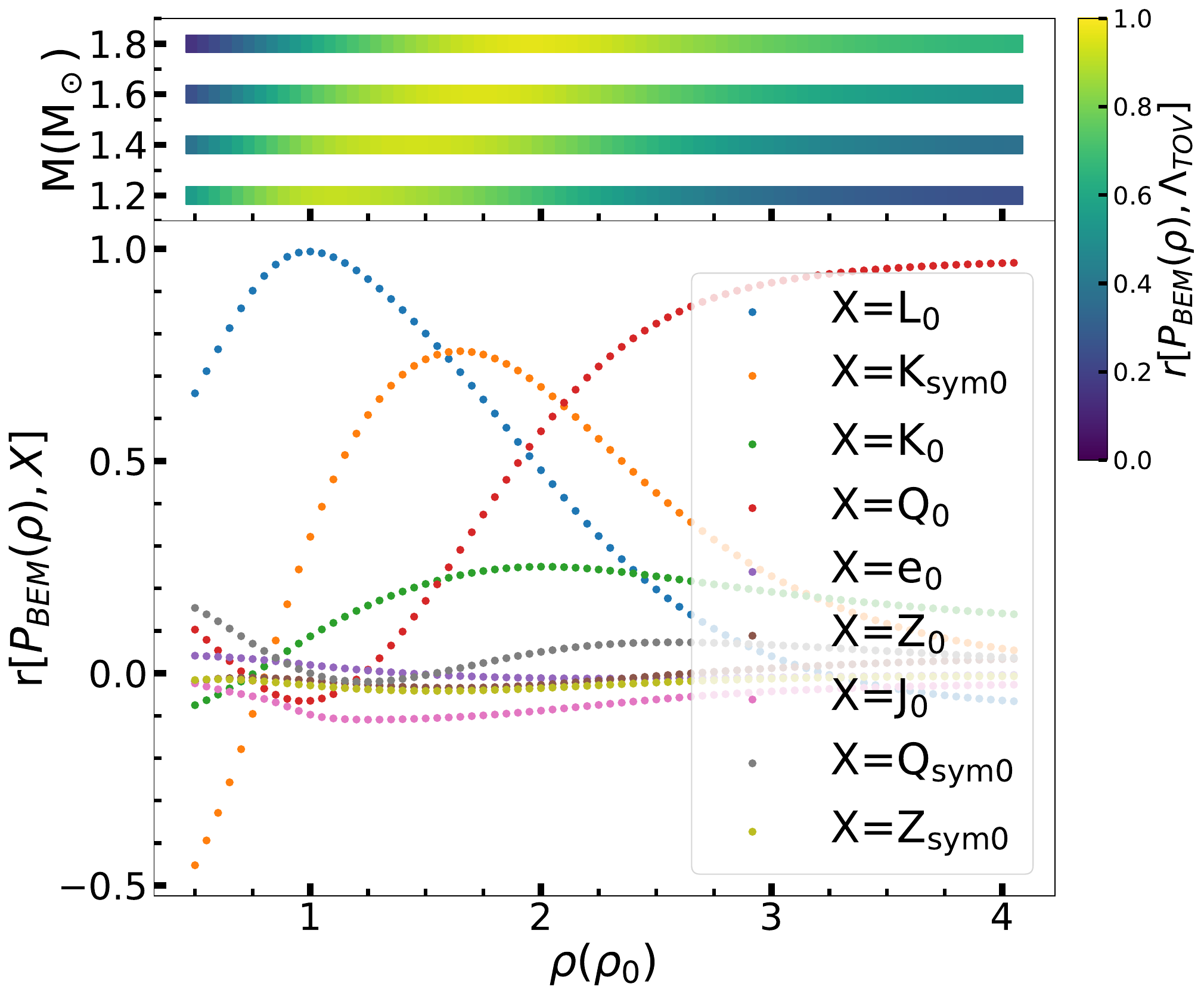}
    \vspace{-0.6cm}
	\caption{\label{fig1}
		Plots for the correlations of  the $\beta$- equilibrium pressure at a given density (P$_{BEM}(\rho)$) with the nuclear matter parameters (lower panel) and with tidal deformability at a given mass $\Lambda_{TOV}(M)$(upper panel). The results are obtained by varying  all the nuclear matter parameters considered (see text for detail). Here it is clear that K$_0$, Q$_{0}$, L$_0$ and K$_{\rm sym0}$ are the most important parameters to model tidal deformability for the NS mass considered.}
\end{figure}

\begin{figure}
	\includegraphics[width=0.5\textwidth]{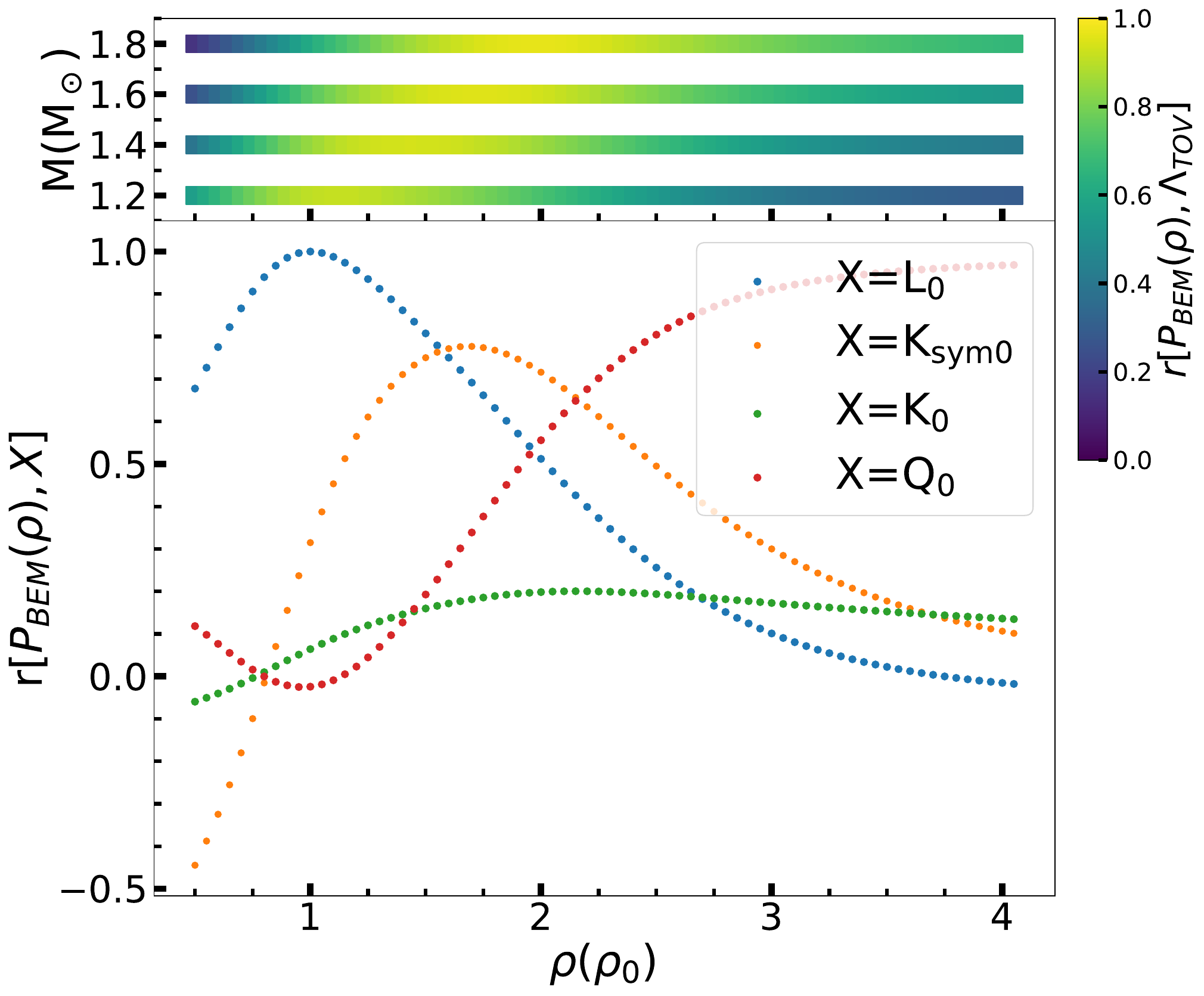}
    \vspace{-0.6cm}
	\caption{Same as Fig.~\ref{fig1}. But, the results are obtained by only varying K$_0$, Q$_{0}$, L$_0$ and K$_{\rm sym0}$ parameters while keeping rest of NMPs fixed at their respective median values.}\label{fig2}
\end{figure}

We aim to identify the subset of the NMPs on which the tidal deformability predominantly depends over a range of NS mass relevant for astrophysical observations of BNS events detectable in the near future. In this context, we study the dependencies of $\Lambda$ on the NMPs in the form of simple polynomial series of those predominant parameters. Such mapping would enable us to evaluate $\Lambda$ without recourse to the computationally expensive solutions of the TOV equations which will pave the way for the Bayesian parameter estimation of the NMPs from the GW events in a computationally efficient method. 

The iso-scalar and iso-vector nuclear matter parameters govern the EoS of $\beta$-equilibrated matter. The iso-scalar NMPs usually considered  to describe the EoS for the SNM are binding energy per nucleon (e$_0$), incompressibility coefficient (K$_0$), skewness (Q$_0$), and kurtosis (Z$_0$). The density dependent symmetry energy that accounts for the deviation from the SNM is governed by the iso-vector NMPs such as symmetry energy coefficient (J$_0$), its slope (L$_0$), curvature (K$_{\rm sym0}$), skewness (Q$_{\rm sym0}$), kurtosis (Z$_{\rm sym0}$) evaluated at $\rho_0$. In order to demonstrate our approach, we use the $\frac{n}{3}$ expansion of the EoS with expansion coefficients depending on the linear combination of the NMPs considered~\cite{Patra:2022yqc}. Only those EoSs are considered which satisfy the condition of (i) thermodynamic stability, (ii) positive semi-definiteness of symmetry energy, (iii) causality of speed of sound, and (iv) maximum mass of the stable non-rotating neutron stars exceeding 2M$_\odot$. A large number of EoSs ($\approx$ 10$^4$) are constructed by drawing all the nine NMPs randomly from their uniform probability distributions (see Table~\ref{tab1}) to compute tidal deformability ($\Lambda_{TOV}$(M)) from the solutions of the TOV equations at a given NS mass M.

{\setlength{\tabcolsep}{1pt}
{\renewcommand{\arraystretch}{1.3}
	\begin{table}[htbp]
		\caption{\label{tab1}Priors for the nuclear matter parameters used in our analysis. All the parameters are uniformly distributed within the minimum (`min') and maximum (`max') bounds. The median (`med') values are also listed \cite{Patra:2022yqc}. All  values are  in the units of MeV.}
	
		\centering
		\begin{ruledtabular}  
			\begin{tabular}{cccccccccc}
				
			 NMP & e$_0$& K$_0$& Q$_0$ & Z$_0$ & J$_0$ & L$_0$ &  K$_{sym0}$ &Q$_{sym0}$ &Z$_{sym0}$\\\cline{1-10}

             min & -16.3 & 200 & -800 & 1400 & 27 & 20 & -250 &300 &-2000\\[1.5ex]

             max & -15.7 & 300 & 800 & 2500 & 37 & 120 & 250 &900 &-1000\\[1.5ex]
            med & -16.0 & 231.96 & -418.89 & 1638.14 & 31.87 & 52.26 & -67.44 &726.49 &-1622.35\\[1.5ex]
				
			\end{tabular}
		\end{ruledtabular}
	\end{table}
In the upper panel of Fig.\ref{fig1}, we display the correlations between $\Lambda_{TOV}$(M) and the pressure of $\beta$-equilibrated matter across various densities spanning from 0.5 - 4$\rho_0$. The density at which maximal correlations emerge exhibits a consistent monotonous increase with the NS mass. Specifically, for a NS with a mass of 1.2 M$_\odot$, the maximum correlation surfaces around 1.25$\rho_0$, while for a NS of 1.8 M$_\odot$, the correlation peak shifts to approximately 2$\rho_0$. This investigation extends to assessing the correlations between the NMPs and the pressure of $\beta$-equilibrated matter at a specific density, depicted in the lower panel of the figure.
The correlations of L$_0$ and K$_{\rm sym0}$ with pressure attain their maximal values at $\rho_0$ and 1.65$\rho_0$, respectively. The correlation between Q$_0$ and pressure increases monotonously beyond $\rho_0$, eventually reaching a saturation point beyond 3$\rho_0$. As for K$_0$, it exhibits its peak correlation with pressure around 2$\rho_0$, albeit significantly less than the values achieved for L$_0$, K$_{\rm sym0}$, and Q$_0$. The remaining NMPs exhibit negligible correlations with pressure, indicated by coefficients $r \leq 0.1$.

To validate our findings, we vary only K$_0$, Q$_0$, L$_0$, and K$_{\rm sym0}$, while keeping rest of the NMPs fixed to their median values  as listed in Table \ref{tab1}. The outcomes are then displayed in Fig.\ref{fig2}. Remarkably, these outcomes closely resemble those of Fig.\ref{fig1}, achieved through variations of all the nine NMPs. This suggests that parameters such as e$_0$, Z$_0$, J$_0$, Q$_{\rm sym0}$, and Z$_{\rm sym0}$ likely exert minimal influence over the values of tidal deformability. These figures play a pivotal role in facilitating the identification of the most pertinent density range for neutron stars of specific masses, intricately tied to the associated nuclear matter parameters. Furthermore, it becomes evident that the values of K$_0$, Q$_0$, L$_0$, and K$_{\rm sym0}$ hold particular significance in shaping the pressure of $\beta$-equilibrated matter within the density range of $\rho$ = 1 - 3$\rho_0$, as depicted in the lower panel of the Figs \ref{fig1} and \ref{fig2}. This observation is particularly pronounced in the strong correlations between these parameters and tidal deformability within the NS mass range of 1.2-1.8 M$_\odot$. Hence, it becomes crucial to parameterize $\Lambda_{TOV}$(M) in terms of these four NMPs, at the very least, rather than considering any subset thereof.
\begin{figure}
	\includegraphics[width=0.45\textwidth]{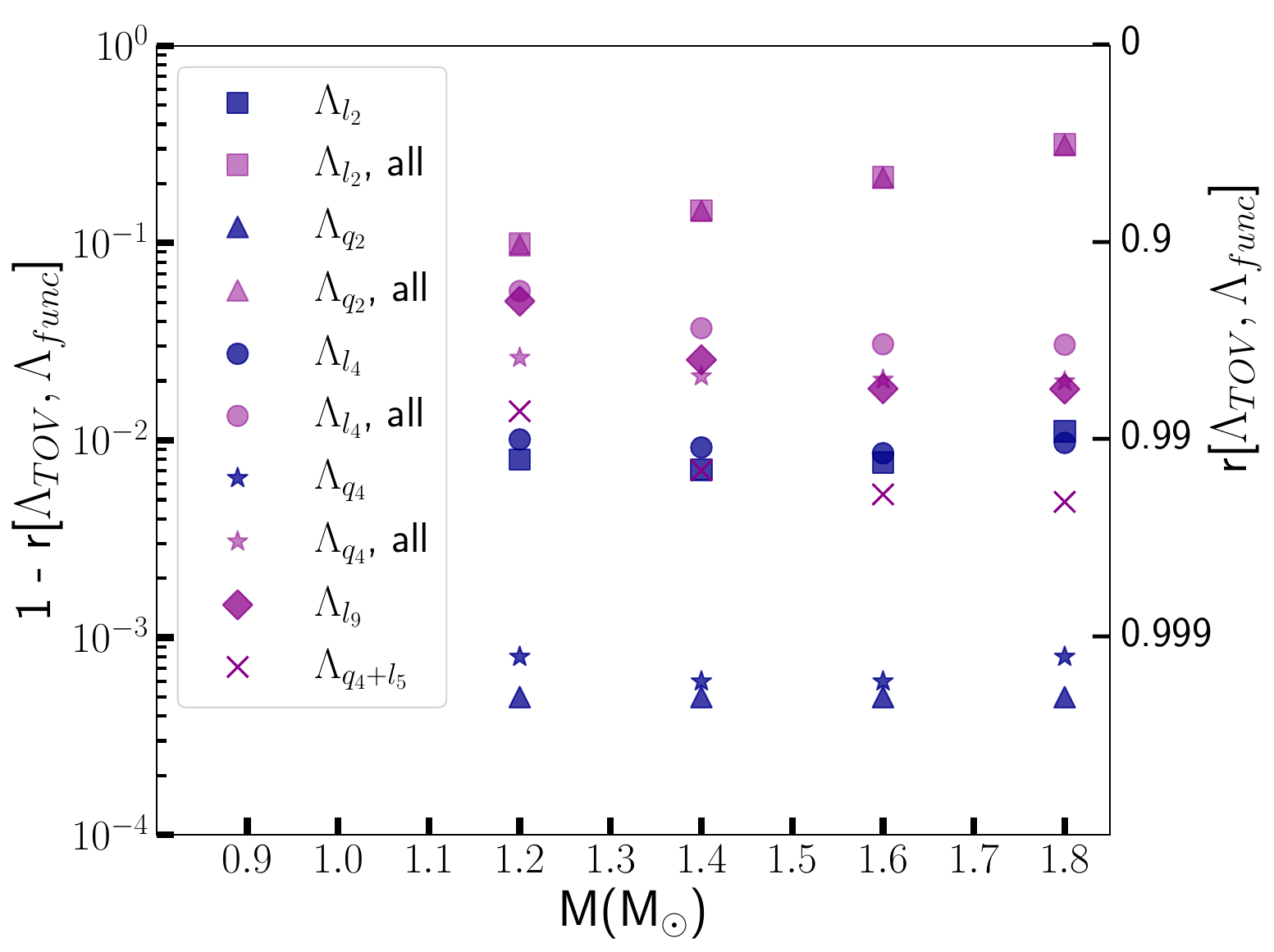}
	\caption{Variations of the correlation coefficients r$[\Lambda_{TOV}$,$\Lambda_{func}]$ as a function of neutron star mass is shown. The $\Lambda_{TOV}$ and $\Lambda_{func}$ are the tidal deformabilities obtained from the solution of TOV equations and those from direct mapping to the different functions of the nuclear matter parameters, respectively. The blue color denotes the the case when only the NMPs involved in the $\Lambda_{func}$ are varied (e.g. $\theta_2$ for $\Lambda_{l_2}$ or $\Lambda_{q_2}$; and $\theta_4$ for $\Lambda_{l_4}$ or $\Lambda_{q_4}$) and the dark magenta color denotes the case $\theta_{all}$ where all the nine NMPs are varied (with an extra label `all') within their respective ranges.}\label{fig3}  
\end{figure}

We express the tidal deformability for a given NS mass using linear ($\Lambda_{l_n}$) and quadratic ($\Lambda_{q_n}$) functions of $n$ number of NMPs as 

\begin{eqnarray}
\Lambda_{l_n} &=& c_0 + \sum_{i=1}^nc_i(x_i-\hat{x}_i) \label{Ln} \\
\Lambda_{q_n} &=& \Lambda_{l_n} + \sum_{i=1}^n\sum_{j=i}^nc_{ij}(x_i-\hat{x}_i)(x_j-\hat{x}_j) \label{Qn}
\end{eqnarray}

where, $x \in$ \{e$_0$, K$_0$, Q$_0$, Z$_0$, J$_0$, L$_0$, K$_{\rm sym0}$, Q$_{\rm sym0}$, Z$_{\rm sym0}$\} for $n = 9$; and $\hat{x}$ corresponds to the median value of parameter $x$. The coefficients c$_i$ and c$_{ij}$ are obtained by fitting the values of $\Lambda_{TOV}$(M) to the Eqs. (\ref{Ln}) and (\ref{Qn}). We consider $\Lambda_{l_n}$, $\Lambda_{q_n}$ with $n = 2$ and $n = 4$ which correspond to $x \in$ \{L$_0$, K$_{\rm sym0}$\} and $x \in$ \{K$_0$, Q$_0$, L$_0$, K$_{\rm sym0}$\}, respectively. We also consider $\Lambda_{l_9}$ which includes all the nine NMPs considered and $\Lambda_{q_4 + l_5}$ with ${q_4}$ denotes contribution up to quadratic order for $x \in$ \{K$_0$, Q$_0$, L$_0$, K$_{\rm sym0}$\} and ${l_5}$ denotes the linear contributions from the remaining 5 NMPs (See Eq. \ref{eqs1}). We refer these fitted functions as $\Lambda_{func}$.

Our general strategy is to first obtain an n-dimensional distribution of the NMPs, keeping all other parameters fixed to their median values \cite{Patra:2022yqc} to compute the NS EoSs. The values of $\Lambda_{TOV}$(M) corresponding to $60\%$ of these EoSs are used to determine the coefficients c$_i$ or c$_{ij}$ in Eqs~\ref{Ln},~\ref{Qn}, and the remaining EoSs are used to assess the merit of the functions. We validate $\Lambda_{func}$ against $\Lambda_{TOV}$ with later one obtained by varying all the NMPs considered uniformly within the  ranges as listed in Table~\ref{tab1}. For convenience, we use the label $\theta_n$  throughout the paper which refers to the variation of a specific set of $n$ number of NMPs. The label $\theta_2$ corresponds to the case where L$_0$ and K$_{\rm sym0}$ are varied, $\theta_4$ represents the variations of K$_0$, Q$_0$, L$_0$, K$_{\rm sym0}$ and $\theta_{all}$ corresponds to the variations of all nine NMPs and $\theta_{2}$ + [`P'] corresponds to the variation of L$_0$, K$_{\rm sym0}$ (i.e. $\theta_2$) together with the variation of the parameter `P', where `P' corresponds to a NMP other than L$_0$ and K$_{\rm sym0}$. ${\theta}_{all}$ - [`P'] denotes the variation of all the parameters except the parameters `P's.

Various $\Lambda_{func}$ are constructed by employing different combinations of NMPs. In Fig~\ref{fig3}, the logarithmic scale is utilized to exhibit the values representing the loss of correlations, quantified as, (1-r[$\Lambda_{TOV}$,  $\Lambda_{func}]$), as a function of NS mass. The values of r[$\Lambda_{TOV}$, $\Lambda_{func}]$ can be read in the linear scale from the ordinate on the right-hand side (major ticks only). Both $\Lambda_{l_2}$, $\Lambda_{q_2}$ functions give the correlation coefficient close to unity with $\Lambda_{TOV}$ for the case of $\theta_2$ (blue symbols) but a substantial loss of correlation is witnessed in the $\theta_{all}$  case (magenta symbols), especially for higher NS masses, $r\sim 0.68$ for M = 1.8 M$_\odot$. Hence, we opt to exclude $\Lambda_{l_2}$, and $\Lambda_{q_2}$ functions from the subsequent analysis. In  the case of $\Lambda_{l_4}$ and $\Lambda_{q_4}$ the correlation coefficient with $\Lambda_{TOV}$ is always close to unity for both $\theta_{4}$ and $\theta_{all}$, almost independent of NS mass. We have also shown the results for $\Lambda_{l_9}$ and $\Lambda_{q_4+l_5}$ functions. These functions exhibit the most robust correlations with $\Lambda_{TOV}$(M), $r\geq 0.95$ for $\Lambda_{l_9}$ and
$r>0.98$ for $\Lambda_{q_4+l_5}$.

\begin{figure}
    \vspace{-1.1cm}
	\includegraphics[width=0.45\textwidth]{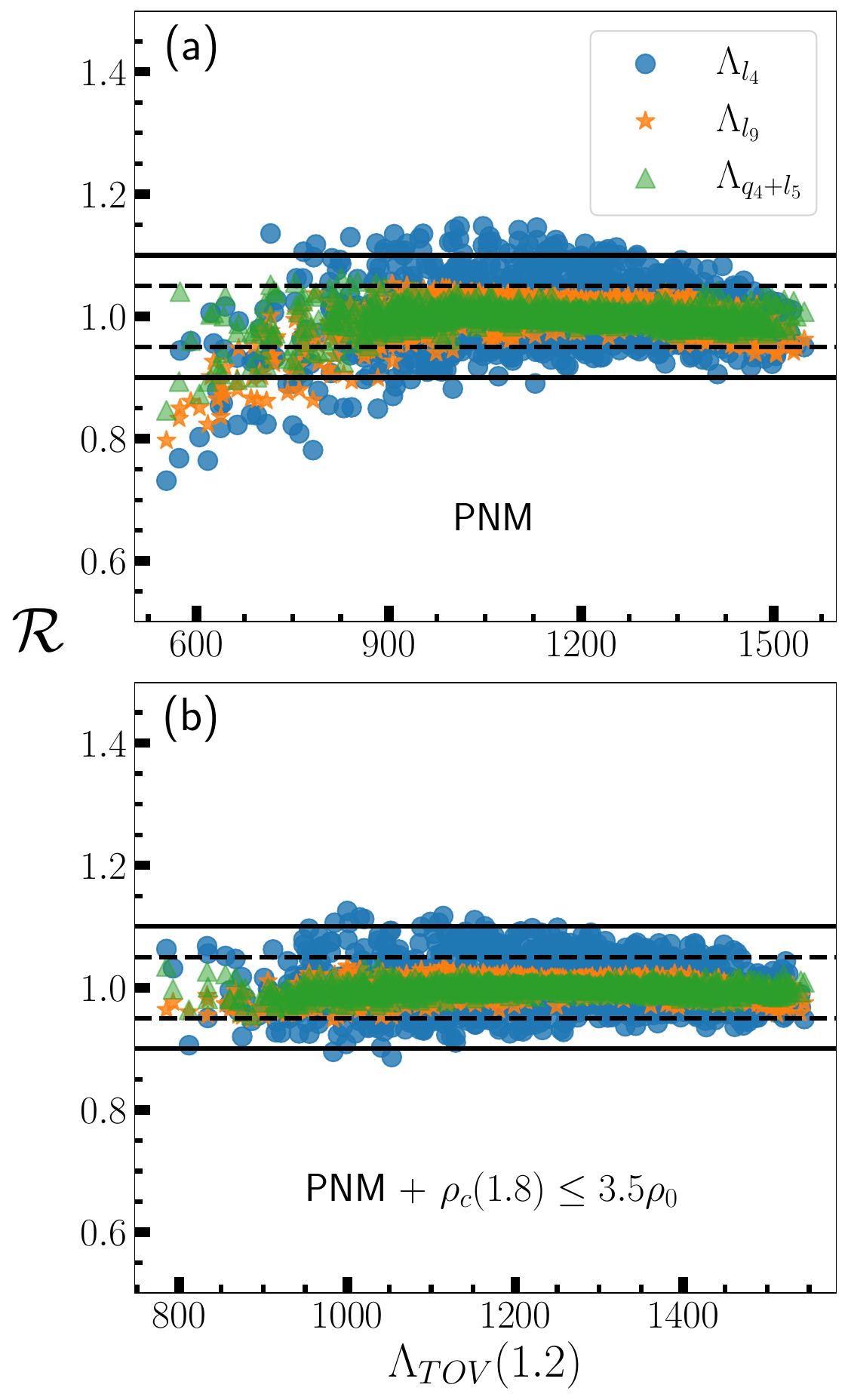}
    \vspace{-0.4cm}
	\caption{The ratio ${\mathcal R}$ = $\frac{\Lambda_{TOV}}{\Lambda{_{func}}}$ for the neutron star mass 1.2 M$_{\odot}$. The $\Lambda_{TOV}$ is obtained by varying all the nuclear matter parameters.The results are for 1000 EoSs randomly drawn from a large sample of EoSs. The upper panel corresponds to the NMP set consistent with energy per particle of PNM within 90$\%$ confidence interval up to 2$\rho_0$ derived from $\chi$EFT. The lower panel has an additional constraint on the central density of 1.8 M$_\odot$ NS ($\rho_c$(1.8)) which is below 3.5$\rho_0$. The dashed and solid horizontal lines represent 5$\%$ and 10$\%$ deviations of $\Lambda_{func}$ from $\Lambda_{TOV}$, respectively.}\label{fig4}
\end{figure}

A detailed systematic analysis has been carried out to find out the most important NMPs. The values of Pearson's correlation coefficient ($r$) between $\Lambda_{TOV}$(M) and  $\Lambda_{l_2}$(M) with M = 1.2 -- 1.8 M$_\odot$ are presented in Table~\ref{tabA1}. Each column represents the variations of the NMPs considered to compute the $\Lambda_{TOV}$(M). The result indicates $\Lambda_{l_2}$, which includes the contributions from  L$_0$ and K$_{\rm sym0}$ only, may not be sufficient to reliably represent $\Lambda_{TOV}$(M) over a wide range of NS mass, the contributions of K$_0$ and Q$_0$ need to be considered (see the discussion of Table~\ref{tabA1} in the Appendix section for details).
\begin{figure}
    \vspace{-1.1cm}
	\includegraphics[width=0.45\textwidth]{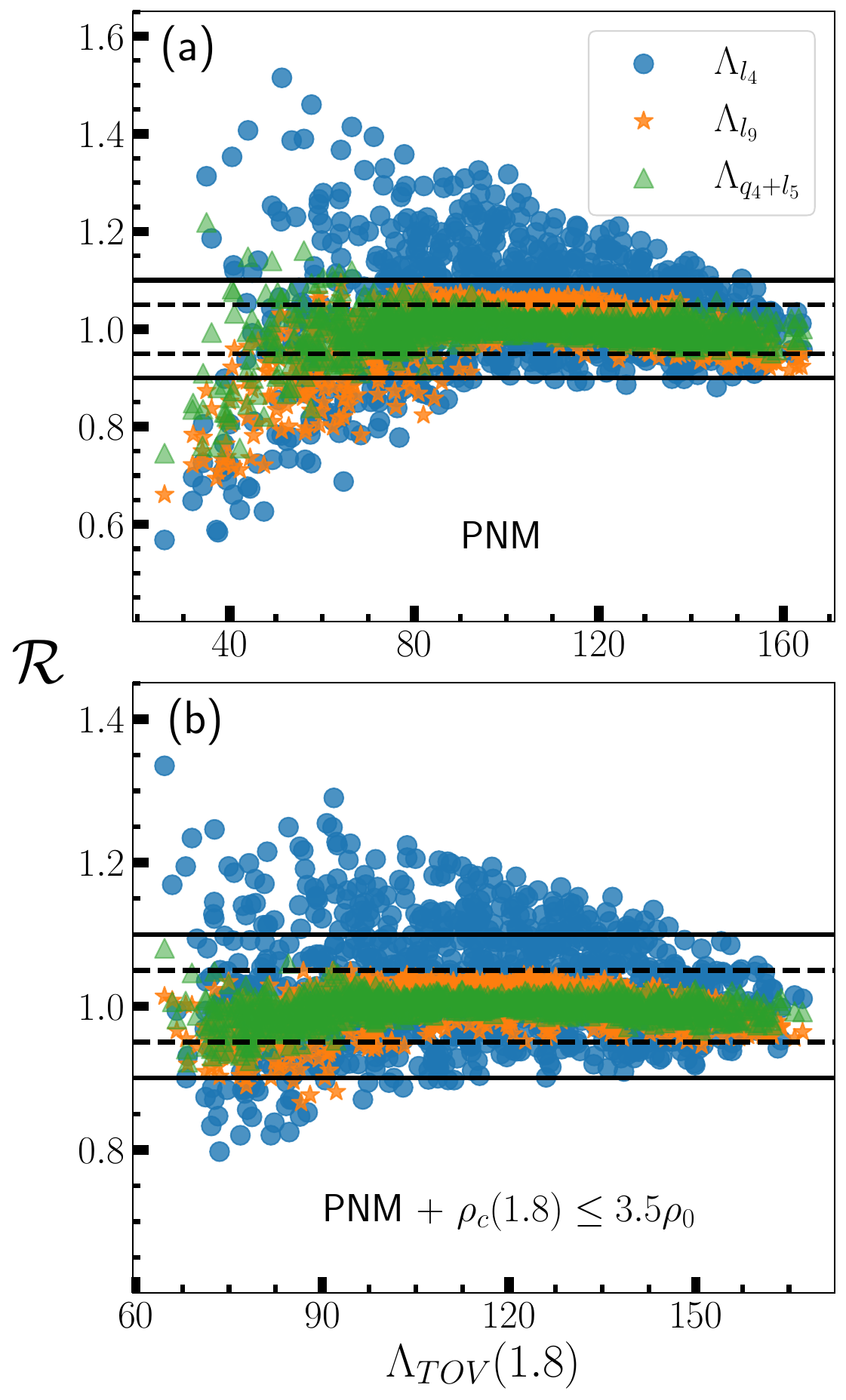}
    \vspace{-0.4cm}
\caption{Same as Fig. \ref{fig4}, but for the neutron star mass 1.8 M$_{\odot}$.}\label{fig5}
\end{figure}

We evaluate the ratio, ${\mathcal R}_M$ = $\frac{\Lambda{_{TOV}}(M)}{\Lambda{_{func}}(M)}$ for the NS masses 1.2 M$_\odot$ - 1.8 M$_\odot$ for three different sets of NMP :
(i) with no constraint
(ii) with pure neutron matter (PNM) constraint 
 (iii) associated with the EoSs for which central baryon density for 1.8 M$_\odot$ ($\rho_{c}$(1.8)) below 3.5$\rho_0$ together with the PNM constraint. In the case of PNM constraint we select those EoSs which satisfy the energy per particle for PNM within 90$\%$ confidence interval up to 2$\rho_0$ derived from $\chi$EFT\cite{Lattimer_rev}. The $\Lambda{_{func}}$  represents $\Lambda_{l_4}$,  $\Lambda_{l_9}$  and $\Lambda_{q_4+l_5}$ for a given NS mass, while  $\Lambda_{TOV}$ values are obtained by varying all the NMPs. The results are plotted  in Figs.\ref{fig4} and \ref{fig5} for NS masses 1.2 M$_\odot$ and 1.8 M$_\odot$, respectively for 1000 randomly selected EoSs from our extensive data set corresponding to the cases (ii) and (iii). The results for $\Lambda_{q_4}$ are quantitatively very close to those for $\Lambda_{l_4}$ and are not shown. Observations from the figures indicate that deviations of the ratio $\mathcal{R}$ from unity are most pronounced for $\Lambda_{l_4}$ and least significant for $\Lambda_{q_4+l_5}$.
 In upper panels, approximately 9$\%$(34$\%$) EoSs for which ratio $\mathcal{R}$ falling outside the range 0.9 to 1.1 for 1.2 (1.8) M$_\odot$ NS for the $\Lambda_{l_4}$ function (see Table \ref{tabA2}). The corresponding deviation for $\Lambda_{q_4+l_5}$ reduces to 0.5$\%$(3.8$\%$) for 1.2 (1.8) M$_\odot$ NS. For the NS with 1.8 M$_\odot$, the EoSs exhibiting a ratio outside the range of 0.9 to 1.1 for $\Lambda_{q_4+l_5}$ function lead to central baryon density, $\rho_c$(1.8)$>$3.5$\rho_0$, requiring the inclusion of higher-order terms in $\Lambda_{func}$. In addition our motivation to restrict the central density of the star always $\leq$ 3.5$\rho_0$ comes from the fact that beyond this density it is very likely to have phase transition to non-nucleonic degrees of freedom e.g. hyperons, quarks, hybrid matter, Bose condensate etc, in which cases a direct comparison between properties of finite nuclei with NS observable becomes inappropriate. Therefore we repeat our calculation by excluding these EoSs and the results are presented in the lower panels. It is clear that now the deviations for $\Lambda_{q_4+l_5}$ from $\Lambda_{TOV}$ for all the EoSs are within 10$\%$.

 \begin{table}[H]
	\caption{\label{tab2}Mean ($\mu$) and standard deviation ($\sigma$) of the ratio ($\mathcal{R}_M = \frac{\Lambda_{TOV}(M)}{\Lambda_{func}(M) }$) for the functions $\Lambda_{l_4}, \Lambda_{q_4}, \Lambda_{l_9}$ and $ \Lambda_{q_4+l_5}$ considered. The values are listed for the NS mass M= 1.2 -- 1.8  M$_\odot $ for the three different NMP sets considered. }
	\centering
	\begin{tabular}{|c|c|cc|cc|cc|cc|}
		\hline   
		& \multirow{4}{*}&\multicolumn{2}{c}{$\Lambda_{l_4}$}\vline &\multicolumn{2}{c}{$\Lambda_{q_4}$}\vline &\multicolumn{2}{c}{$\Lambda_{l_9}$}\vline &\multicolumn{2}{c}{$\Lambda_{q_4+l_5}$}\vline\\  \cline{3-10}
		Constraints    & Ratio & $\mu$  & $\sigma$ & $\mu$ & $\sigma$  & $\mu$ & $\sigma$ & $\mu$ & $\sigma$ \\ \cline{1-10}
		& $\mathcal{R}_{\rm 1.2}$ & 1.01 & 0.12 & 1.01 & 0.07 & 1.01 & 0.12 & 1.00 & 0.06  \\[1.5ex]  \cline{2-10}
		\hspace{0.1cm}Without PNM & $\mathcal{R}_{\rm 1.4}$ & 1.01 & 0.09 & 1.01& 0.07 & 1.00 & 0.08 & 1.00 & 0.04    \\[1.5ex]  \cline{2-10}
		& $\mathcal{R}_{\rm 1.6}$ & 1.01 & 0.08 & 1.01 & 0.07 & 1.00 & 0.07 & 1.00 & 0.04  \\[1.5ex]  \cline{2-10}
		& $\mathcal{R}_{\rm 1.8}$ & 1.02 & 0.10 & 1.02 & 0.09 & 1.00 & 0.07 & 1.00 & 0.04  \\[1.5ex]  \cline{1-10}

        & $\mathcal{R}_{\rm 1.2}$ & 1.01 & 0.06 & 1.01 & 0.07 & 1.00 & 0.03 & 1.00 & 0.02  \\[1.5ex]  \cline{2-10}
		\hspace{0.1cm}With PNM & $\mathcal{R}_{\rm 1.4}$ & 1.01 & 0.07 & 1.02 & 0.09 & 1.00 & 0.04 & 1.00 & 0.02    \\[1.5ex]  \cline{2-10}
		& $\mathcal{R}_{\rm 1.6}$ & 1.02 & 0.09 & 1.03 & 0.12 & 1.00 & 0.05 & 1.00 & 0.03  \\[1.5ex]  \cline{2-10}
		& $\mathcal{R}_{\rm 1.8}$ & 1.03 & 0.12 & 1.06 & 0.22 & 1.00 & 0.06 & 1.00 & 0.04  \\[1.5ex]  \cline{1-10}
		
		& $\mathcal{R}_{\rm 1.2}$ & 1.01 & 0.04 & 1.01 & 0.05 & 1.00 & 0.01 & 1.00 & 0.01  \\[1.5ex]  \cline{2-10}
		\hspace{0.1cm}With PNM
  & $\mathcal{R}_{\rm 1.4}$ & 1.01 & 0.05 & 1.01 & 0.06 & 1.00 & 0.02 & 1.00 & 0.01    \\[1.5ex]  \cline{2-10}
 + & $\mathcal{R}_{\rm 1.6}$ & 1.02 & 0.06 & 1.02 & 0.07 & 1.00 & 0.02 & 1.00 & 0.01  \\[1.5ex]  \cline{2-10}
	$\rho_{c}$(1.8)$\leq$3.5$\rho_0$ & $\mathcal{R}_{\rm 1.8}$ & 1.02 & 0.08 & 1.03 & 0.09 & 1.00 & 0.03 & 1.00 & 0.02  \\[1.5ex]  \cline{1-10}
				\end{tabular}
\end{table}

 In Table~\ref{tab2}, the mean and the standard deviation are presented for the ratio $ \mathcal{R}_M = \frac{\Lambda_{TOV}(M)}{\Lambda_{func}(M)}$ for the NS mass in the range of $1.2 - 1.8M_\odot$ for the functions $\Lambda_{l_4}$, $\Lambda_{q_4}$, $\Lambda_{l_9}$, as defined by Eq.~(\ref{Ln}) and Eq.~(\ref{Qn}) for the three different NMP sets considered. The results are also presented for the  mixed function $\Lambda_{q_4+l_5}$ which is expressed as in Eq\ref{eqs1}. It is clear that the inclusion of the constraint from low density PNM EoS have significantly improved the agreement between the values of $\Lambda_{func}$ with $\Lambda_{TOV}$. The mean values of the
ratio with the inclusion of the constraint are close to unity for all the NS masses for $\Lambda_{l_4}$ and $\Lambda_{q_4}$ functions. The inclusions of more number of parameters further improves the values of
standard deviation. For example, in the case of $\Lambda_{q_4+l_5}$ ($\Lambda_{l_9}$) function, the standard deviation of the ratio between $\Lambda_{func}$ and $\Lambda_{TOV}$ for 1.2 M$_\odot$ NS is reduced to 0.02 (0.03) which is 0.06 for $\Lambda_{l_4}$ function. The results improved significantly if the softest EoSs having central density 3.5$\rho_0$ correspond to 1.8 M$_\odot$ NS are removed.
 \begin{figure}[H]
	\includegraphics[width=0.5\textwidth]{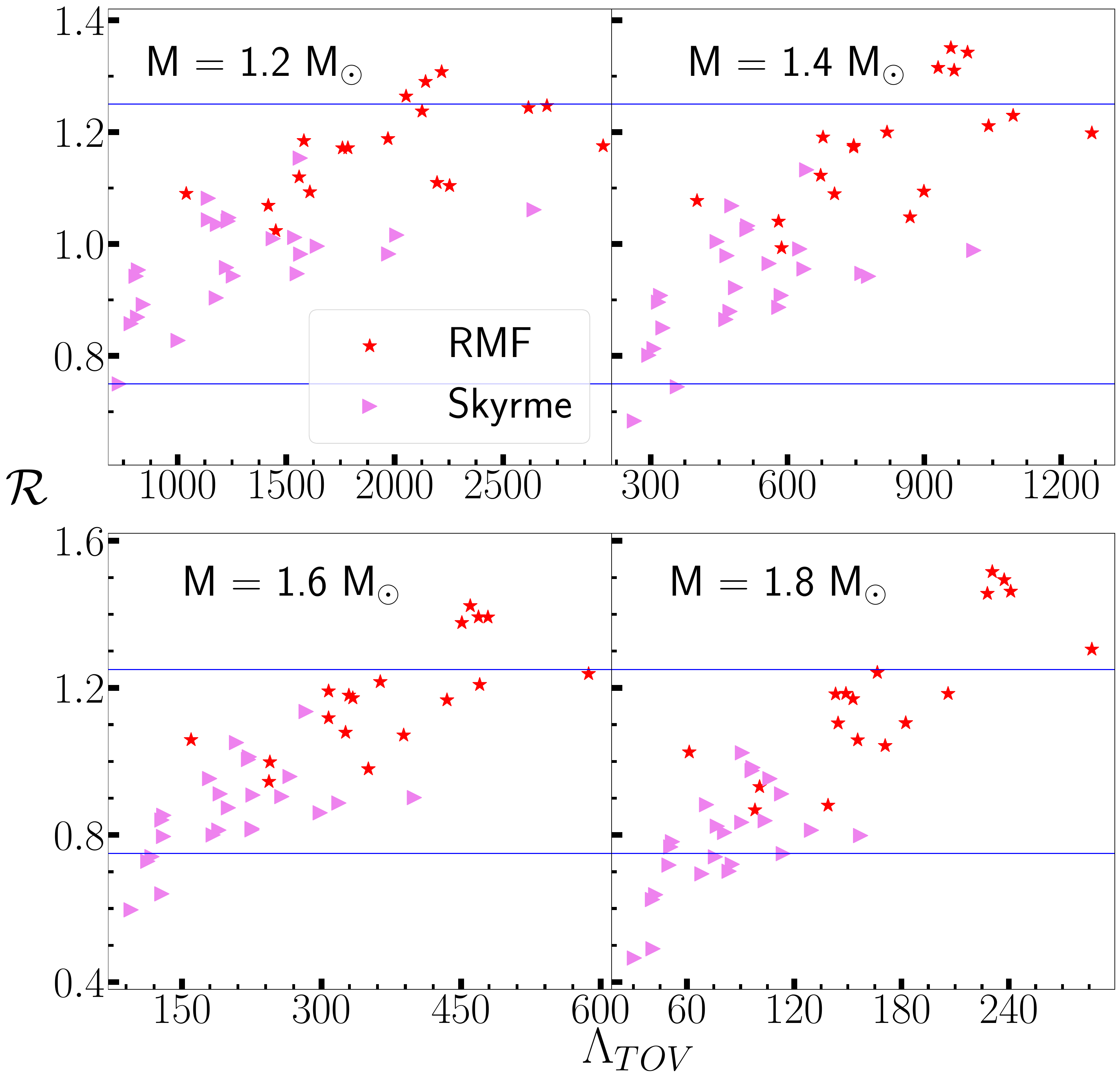}
	\caption{Predictions of  $\Lambda_{l_4}$ function, fitted with the PNM constraint from $\chi$EFT, for 24 non-relativistic models (triangles) and 18 relativistic models (asterisks). The horizontal lines shows 25$\%$ deviation from the TOV value.}\label{fig6}
\end{figure}
We also compare our $\Lambda_{func}$ against $\Lambda_{TOV }$ obtained using the EoSs from non-relativistic and relativistic mean-field models. With the existing NMPs available in the literature, we can only compute $\Lambda_{l_4}$ for these mean-field models. The results for the ratio ${\mathcal R}$ are presented in Fig~\ref{fig6}. Most of the values fall within $25\%$ in comparison to their actual $\Lambda_{TOV}$ values. The number of points having larger deviations increase with the NS mass. Such deviations may not be very surprising, since, the function $\Lambda_{l_4}$ is fitted to the EoSs for the $\frac{n}{3}$ model which might have different behavior for the EoSs compared to the considered MFMs. It's worth mentioning that the utilization $\Lambda_{q_4 + l_5}$ may improve the accuracy of these predictions.

Finally, in Tables~\ref{tabA4}-~\ref{tabA6} of the Appendix the fitted coefficients of $\Lambda_{l_4}$, $\Lambda_{l_9}$ and $\Lambda_{q_4+l_5}$ functions are listed, respectively, for the NS masses considered. They may be employed to  estimate quickly the values of tidal deformability  without recourse to the solution of TOV equations. Our proposed functions for the tidal deformability would facilitate the Bayesian analysis, which often entail the calculation of tidal deformability across a broad spectrum of NS masses for a substantial number of EoSs ($\sim 10^6$). 

\section{Summary}\label{summ}

We performed an extensive analysis aimed at identifying the key nuclear matter parameters that primarily influence the tidal deformability values of neutron stars. Among these parameters, both the iso-scalar parameters K$_0$ and Q$_0$, as well as the iso-vector parameters L$_0$ and K$_{\rm sym0}$, emerged as the most significant contributors. We fit the values of tidal deformabilities obtained by solving the TOV equations to the linear and quadratic functions of these nuclear matter parameters. To validate these functions, we compared them with the tidal deformability values obtained from TOV equations applied to equations of state constructed with varying all the nuclear matter parameters. Our analysis showed that the predictions for $\Lambda_{q_4+l_5}$ deviate within 10$\%$ from the $\Lambda_{TOV}$ for EoSs constrained by the N$^3$LO $\chi$EFT. We established a direct mapping between tidal deformability values and nuclear matter parameters, enabling quick estimations without recourse to the solution of TOV equations. Importantly, this approach can be extended to various other EoS models and neutron star observables. Consequently, it will facilitate efficient Bayesian statistical inference of relevant nuclear matter parameters directly from astrophysical observations. The efforts to further improve  the relation of tidal deformability with the nuclear matter parameters are underway.

\section{Acknowledgements} 
\vspace{-0.12cm}
AM acknowledges support from the DST-SERB Start-up Research Grant SRG/2020/001290. BKA acknowledges partial support from the SERB, Department of Science and Technology, Government of India with  CRG/2021/000101.  

\bibliographystyle{apsrev4-2}
\pagebreak
\setcounter{equation}{0}
\makeatletter
\renewcommand{\theequation}{A\arabic{equation}}

\widetext
\section*{Appendix}

\begin{table}[H]
		\caption{\label{tabA1}The Pearson's correlation coefficient, `$r$' between tidal deformabilities $\Lambda_{l_2} (M)$ and  $\Lambda_{TOV} (M)$ for a given neutron star mass $M (M_\odot)$. The values of $\Lambda_{TOV}$ employed to obtain the results presented in 4th to 10th columns involve variations of the specified nuclear matter parameter in addition to $\theta_2$ which represents the variations of L$_0$ and K$_{sym0}$ only. The last column corresponds to the variation of all the nuclear matter parameters except K$_0$ and Q$_0$ which is denoted as $\theta_{all}$-[K$_0$,Q$_0$].}
		\centering
		\begin{ruledtabular}  
			\begin{tabular}{ccccccccccc}
                M(M$_\odot$)& $\theta_2$ & {$\theta_{all}$} & {$\theta_2$+[e$_0$]} & {$\theta_2$+[K$_0$]} & {$\theta_2$+[Q$_0$]} & {$\theta_2$+[Z$_0$]} & {$\theta_2$+[J$_0$]} & {$\theta_2$+[Q$_{sym0}$]} & {$\theta_2$+[Z$_{sym0}$]} &  $\theta_{all}$-[K$_0$,Q$_0$] \\ [1.5ex] \hline 
           1.2  & 0.97 & 0.90 & 0.99 & 0.97 & 0.90 & 0.97 & 0.96 & 0.97 & 0.97 & 0.95 \\ [1.5ex] \hline

          1.4  & 0.99 & 0.88 & 1.00 & 0.96 & 0.88 & 0.99 & 0.97 & 0.98 & 0.99 & 0.96 \\ [1.5ex] \hline

           1.6  & 0.99 & 0.78 & 1.00 & 0.94 & 0.83 & 0.99 & 0.98 & 0.97 & 0.99 & 0.95 \\ [1.5ex] \hline

           1.8 & 0.99 & 0.68 & 1.00 & 0.86 & 0.73 & 0.98 & 0.97 & 0.95 & 0.99 & 0.93 \\ [1.5ex] \hline
			\end{tabular}
		\end{ruledtabular}
	\end{table}
In Table~\ref{tabA1} we present the values  of correlation between $\Lambda_{l_2}$ and $\Lambda_{TOV}$. The correlations are close to unity for the case of  $\theta_2$ irrespective of  NS mass considered. The correlation decreases with an increase in NS mass for the case of $\theta_{all}$ which includes the variations of additional parameters, not considered in the fit of the function $\Lambda_{l_2}$ indicating that n=2 in Eq.~(\ref{Ln}) may not adequately determine the values of tidal deformability for higher NS masses. So there are some other parameters which need to be included in the Eq.~(\ref{Ln}) to improve the representation of $\Lambda_{TOV}$. To identify the additional NMPs which need to be included  we consider now the case of $\theta_{2}$ + [`P']. The ${\theta}_{all}$ - [$K_0,Q_0$] in the last column denotes the result obtained by varying all the NMPs except  the $K_0$ and $Q_0$. It may be inferred from the 3rd and last column of the table that the K$_0$ and Q$_0$ are also important to determine the values of tidal deformability. This is also evident from the 5th and 6th columns. The remaining 5 parameters e$_0$, Z$_0$, J$_0$, Q$_{sym0}$, Z$_{sym0}$ do not seem to contribute significantly to the values of tidal deformability. These trends re-insure K$_0$, Q$_0$, L$_0$ and K$_{sym0}$ have a greater impact on the tidal deformability. The trends are similar for correlations of $\Lambda_{TOV}(M)$ with $\Lambda_{q_2}(M)$ and are not presented here.

\begin{table}[H]
	\caption{Percentage of outliers for different models depending on the criteria given in `criteria' column for the three different NMP sets considered. Here N$_{\mathcal R_{M}}$(in $\%$) are the percentage of outliers for the ratio ${\mathcal R_{M}}=\frac{\Lambda_{TOV}}{\Lambda_{func}}$ for a given mass M for a given function $\Lambda_{func}$. }
	\label{tabA2}
	\centering
	\begin{tabular}{|c|c|cc|cc|cc|cc|}
		\hline   
		& \multirow{4}{*}&\multicolumn{2}{c}{$\Lambda_{l_4}$}\vline &\multicolumn{2}{c}{$\Lambda_{q_4}$}\vline &\multicolumn{2}{c}{$\Lambda_{l_9}$}\vline &\multicolumn{2}{c}{$\Lambda_{q_4+l_5}$}\vline\\  \cline{3-10}
		Constraints    & criteria & N$_{\mathcal R_{1.2}}$  & N$_{\mathcal R_{1.8}}$ & N$_{\mathcal R_{1.2}}$  & N$_{\mathcal R_{1.8}}$  &  N$_{\mathcal R_{1.2}}$  & N$_{\mathcal R_{1.8}}$  & N$_{\mathcal R_{1.2}}$  & N$_{\mathcal R_{1.8}}$ \\ \cline{1-10}
        & 0.9$\geq\mathcal{R}\geq$1.1 & 32 & 24.8 & 11.9 & 16.5 & 26.9 & 12.5 & 7.6 & 3.3 \\[1.5ex]  \cline{2-10}
		\hspace{0.1cm}Without PNM & 0.7$\geq\mathcal{R}\geq$1.3 & 2.8 & 1.1 & 0.1 & 1.4 & 2.9 & 0.5 & 0.2 & 0.2  \\[1.5ex]  \cline{1-10}

		& 0.9$\geq\mathcal{R}\geq$1.1 & 9.1 & 34.5 & 11.2 & 33.2 & 2.7 & 6.9 & 0.5 & 3.8 \\[1.5ex]  \cline{2-10}
		\hspace{0.1cm}With PNM & 0.7$\geq\mathcal{R}\geq$1.3 & 0 & 3.5 &  0.3 & 6.6 & 0 & 1 & 0 & 0.1  \\[1.5ex]  \cline{1-10}

		& 0.95$\geq\mathcal{R}\geq$1.05 & 24.4 & 50.4 & 25.5 & 51.9 & 0.3 & 7.2 & 0 & 1.9 \\[1.5ex]  \cline{2-10}
		\hspace{0.1cm}With PNM + $\rho_{c}$(1.8)$\leq$3.5$\rho_0$ & 0.9$\geq\mathcal{R}\geq$1.1& 1.1 & 21.5 & 3.5 & 20.2 & 0 & 0.5 & 0 & 0  \\[1.5ex]  \cline{1-10}

				\end{tabular}
\end{table}

\bea
\Lambda_{q_4+l_5} &=& c_0 + c_1(K_0 - \bar{K}_0) + c_2(Q_0 - \bar{Q}_0) + c_3(L_0 - \bar{L}_0) + c_4(K_{sym0} - \bar{K}_{sym0}) + c_{11}(K_0 - \bar{K}_0)^2 \nonu \\
 &+& c_{12}(K_0 - \bar{K}_0)(Q_0 - \bar{Q}_0) + c_{13}(K_0 - \bar{K}_0)(L_0 - \bar{L}_0) + 
c_{14}(K_0 - \bar{K}_0)(K_{sym0} - \bar{K}_{sym0}) \nonu \\ 
& + &  c_{22}(Q_0 - \bar{Q}_0)^2 + c_{23}(Q_0 - \bar{Q}_0)(L_0 - \bar{L}_0) + 
c_{24}(Q_0 - \bar{Q}_0)(K_{sym0} - \bar{K}_{sym0}) +
c_{33}(L_0 - \bar{L}_0)^2 \nonu \\
& + & c_{34}(L_0 - \bar{L}_0)(K_{sym0} - \bar{K}_{sym0}) + c_{44}(K_{sym0} - \bar{K}_{sym0})^2 +  c_{1L}(e_0 - \bar{e}_0) + c_{2L}(Z_0 - \bar{Z}_0) \nonu \\
& + & c_{3L}(J_0 - \bar{J}_0) + c_{4L}(Q_{sym0} - \bar{Q}_{sym0}) + c_{5L}(Z_{sym0} - \bar{Z}_{sym0}).\label{eqs1}
\eea
It may be noted that the nuclear matter parameters $K_0, Q_0, L_0, K_{\rm sym,0}$  are considered up to the quadratic order, whereas, for the remaining parameters $e_0, Z_0, J_0, Q_{\rm sym,0}$ and $Z_{\rm{sym},0}$ only the contributions up to the linear order are included. The mean values of the ratio with the inclusion of the constraints are close to unity for all the NS masses for $\Lambda_{l_4}$ and $\Lambda_{q_4}$ functions. The results improved significantly in the case of $\Lambda_{q_4+l_5}$ and $\Lambda_{l_9}$ functions. For example, in the case of $\Lambda_{q_4+l_5}$ ($\Lambda_{l_9}$) function, the standard deviation of the ratio between $\Lambda_{func}$ and $\Lambda_{TOV}$ for 1.2 M$_\odot$ NS is reduced to 0.02 (0.04) which is 0.06 for $\Lambda_{l_4}$ function in the case of PNM constraint.

In Table~\ref{tabA2} we have shown the percentage of outliers (N$_{\mathcal R_{M}}$)(in $\%$) for the NS of mass M (M$_\odot$), which are deviating from the respective $\Lambda_{TOV}$ values by 5$\%$, 10$\%$ and 30$\%$ for the three different NMP sets considered. We found $\Lambda_{q_4+l_5}$ to be the best-fit function.
{\renewcommand{\arraystretch}{1.3}
 \begin{table}[htbp]
		\caption{The fitted  values of coefficients `Coef'  appearing in  Eq. (\ref{Ln})  for the  $\Lambda_{l_4}$ function for the NS mass 1.2 -- 1.8 M$_\odot$ for two different NMP sets : first, utilizing low density PNM data from N$^{3}$LO $\chi$EFT and second, incorporating an additional constraint on central baryon density of 1.8 M$_\odot$ NS (see text for detail).}
		\label{tabA4}
		\centering
			\begin{tabular}{|c|c|c|c|c|}
				\hline
    &\multicolumn{4}{c}{With PNM}\vline
				\\ [1.5ex] \cline{2-5}
                 &\multicolumn{4}{c}{Mass(M$_\odot$)}\vline
				\\ [1.5ex] \cline{2-5}
			     Coef  & 1.2  &  1.4  &  1.6  & 1.8  \\ [1.5ex] \cline{1-5} 
				c$_0$ & 1092.89 &  445.99 & 190.59 &  81.84 \\ [1.5ex] \cline{1-5} 
                c$_1$ & 2.84 & 1.37 &   0.70 & 0.37 \\ [1.5ex] \cline{1-5} 
                c$_2$ & 0.44 & 0.23 &  0.13 & 0.07 \\ [1.5ex] \cline{1-5} 
                c$_3$ & 10.47 & 3.79 &    1.45 & 0.56  \\ [1.5ex] \cline{1-5} 
                c$_4$ &  2.16 &    1.02 & 0.50 & 0.25\\ [1.5ex] \cline{1-5}	
			\end{tabular}
\begin{tabular}
{|c|c|c|c|}
				\hline
    \multicolumn{4}{|c|}{With PNM + $\rho_{c}$(1.8)$\leq$3.5$\rho_0$}\vline
				\\ [1.5ex] \cline{1-4}
                 \multicolumn{4}{|c|}{Mass(M$_\odot$)}\vline
				\\ [1.5ex] \cline{1-4}
			      1.2  &  1.4  &  1.6  & 1.8  \\ [1.5ex] \cline{1-4} 
				 1118.76 & 459.55 & 198.01 & 86.11 \\ [1.5ex] \cline{1-4} 
                 2.67  & 1.30  &  0.66 & 0.35 \\ [1.5ex] \cline{1-4} 
                 0.39 & 0.21  &  0.12  & 0.07 \\ [1.5ex] \cline{1-4} 
                 10.63  &  3.90  &  1.53  & 0.62  \\ [1.5ex] \cline{1-4} 
                 1.94  &  0.93  &  0.46 &  0.23 \\ [1.5ex] \cline{1-4} 
\end{tabular}
	\end{table}

{\renewcommand{\arraystretch}{1.3}
	\begin{table}
		\caption{ Same as Table \ref{tabA4} but for $\Lambda_{l_9}$ function.}
		\label{tabA5}
		\centering
			\begin{tabular}{|c|c|c|c|c|}
				\hline   
				&\multicolumn{4}{c}{With PNM}\vline
				\\ [1.5ex] \cline{2-5}
                 &\multicolumn{4}{c}{Mass(M$_\odot$)}\vline
				\\ [1.5ex] \cline{2-5}
			     Coef  & 1.2  &  1.4  &  1.6  & 1.8  \\ [1.5ex] \cline{1-5} 
				c$_0$ & 1092.02 & 444.79 & 190.08 & 81.88  \\ [1.5ex] \cline{1-5} 
                c$_1$ & -1.74 & -1.04 &   -0.64 & -0.41   \\ [1.5ex] \cline{1-5} 
                c$_2$ & 2.95 & 1.41 &    0.71 & 0.37  \\ [1.5ex] \cline{1-5} 
                c$_3$ & 0.44 & 0.23 &   0.13 & 0.07  \\ [1.5ex] \cline{1-5} 
                c$_4$ & 0.02 & 0.01 &   0.008 & 0.005  \\ [1.5ex] \cline{1-5}
                c$_5$ & -3.52 & -1.57 &  -0.83 & -0.51 \\ [1.5ex] \cline{1-5}
                c$_6$ & 10.96 & 3.98 &    1.55 & 0.62\\ [1.5ex] \cline{1-5}
                c$_7$ & 2.14 & 1.01 &    0.49 & 0.24 \\ [1.5ex] \cline{1-5}
                c$_8$ & 0.33 & 0.17 &    0.09 & 0.05 \\ [1.5ex] \cline{1-5}
                c$_9$ & 0.02 & 0.009 &    0.006 &  0.003 \\ [1.5ex] \cline{1-5}
                \end{tabular}
            \begin{tabular}
            {|c|c|c|c|}
				\hline  
			\multicolumn{4}{|c|}{With PNM + $\rho_{c}$(1.8)$\leq$3.5$\rho_0$}\vline
				\\ [1.5ex] \cline{1-4}
                 \multicolumn{4}{|c|}{Mass(M$_\odot$)}\vline
				\\ [1.5ex] \cline{1-4}
			     1.2  &  1.4  &  1.6  & 1.8  \\ [1.5ex] \cline{1-4} 
				1118.08 & 459.32 & 198.45  & 86.90  \\ [1.5ex] \cline{1-4} 
                -2.50   & -1.34  &  -0.74  & -0.43   \\ [1.5ex] \cline{1-4} 
                 2.66  & 1.28 &  0.65 & 0.34  \\ [1.5ex] \cline{1-4} 
               0.39  &  0.21 &  0.12  & 0.08  \\ [1.5ex] \cline{1-4} 
                0.02 &  0.01  &  0.01 &  0.005   \\ [1.5ex] \cline{1-4}
                 -5.75 &  -2.58 &  -1.29  &  -0.70 \\ [1.5ex] \cline{1-4}
                 11.05 &  4.07 &  1.61 & 0.66  \\ [1.5ex] \cline{1-4}
                1.89 &  0.90 &  0.45 &  0.22 \\ [1.5ex] \cline{1-4}
               0.28 &  0.14 &  0.08  &  0.04 \\ [1.5ex] \cline{1-4}
                0.012   &  0.007  &  0.004 &  0.003 \\ [1.5ex] \cline{1-4} 
				
			\end{tabular}
	\end{table}
		
{\renewcommand{\arraystretch}{1.3}
	\begin{table}[htbp]
		\caption{Same as Table \ref{tabA4} but for $\Lambda_{q_4 + l_5}$ function. }\label{tabA6}
 		\centering
			\begin{tabular}{|c|c|c|c|c|}
				\hline   
               &\multicolumn{4}{c}{With PNM}\vline
				\\ [1.5ex] \cline{2-5}
				&\multicolumn{4}{c}{Mass, M (M$_\odot$)}\vline
				\\ [1.5ex] \cline{2-5} 
			     Coef  & 1.2  &  1.4  &  1.6  & 1.8  \\ [1.5ex] \cline{1-5} 
                c$_0$ & 1093.83 &  442.79 & 186.85 & 78.36 \\ [1.5ex] \cline{1-5} 
                c$_1$ & 3.44 & 1.70 &   0.87 & 0.46 \\ [1.5ex] \cline{1-5} 
                c$_2$ & 0.52 & 0.28 &   0.16 & 0.01  \\ [1.5ex] \cline{1-5} 
                c$_3$ &9.98 & 3.35 &    1.15 & 0.38   \\ [1.5ex] \cline{1-5} 
                c$_4$ & 2.15 & 1.0 & 0.47 &   0.21\\ [1.5ex] \cline{1-5} 
                c$_{11}$  & -$5.44 \times 10^{-3}$ &  -$1.88 \times 10^{-3}$ &  -$5.94 \times 10^{-4}$ & -$1.13 \times 10^{-4}$ \\ [1.5ex] \cline{1-5} 
                c$_{12}$ & -$1.70 \times 10^{-3}$ & -$7.88 \times 10^{-4}$ &  -$3.74 \times 10^{-4}$ & -$1.76 \times 10^{-4}$   \\ [1.5ex] \cline{1-5} 
                c$_{13}$ & 0.04  &   0.02 & $9.29 \times 10^{-3}$  &  $4.26 \times 10^{-3}$\\ [1.5ex] \cline{1-5}
                c$_{14}$ & -$7.99 \times 10^{-3}$ &   -$2.78 \times 10^{-3}$ &  -$8.98 \times 10^{-4}$ & -$2.07 \times 10^{-4}$\\ [1.5ex] \cline{1-5} 
                c$_{22}$ & -$1.16 \times 10^{-4}$  & -$6.5 \times 10^{-5}$ &  -$3.8 \times 10^{-5}$ & -$2.4 \times 10^{-5}$ \\ [1.5ex] \cline{1-5} 
                c$_{23}$ &  $4.27 \times 10^{-3}$ &  $2.43 \times 10^{-3}$ &   $1.40 \times 10^{-3}$ &  $7.89 \times 10^{-4}$ \\ [1.5ex] \cline{1-5} 
                c$_{24}$ & -$1.18 \times 10^{-3}$ &  -$4.87 \times 10^{-4}$ & -$1.87 \times 10^{-4}$ & -$5 \times 10^{-5}$\\ [1.5ex] \cline{1-5}
                c$_{33}$ & 0.05  &  0.02  &  $5.55 \times 10^{-3}$ & $2.69 \times 10^{-3}$ \\ [1.5ex] \cline{1-5} 
                c$_{34}$ &  0.01 &  $7.12
                \times 10^{-3}$ &  $3.59 \times 10^{-3}$ &  $1.82 \times 10^{-3}$  \\ [1.5ex] \cline{1-5}
                c$_{44}$ & -$4.55 \times 10^{-3}$ &  -$2.05 \times 10^{-3}$  &   -$1.0 \times 10^{-3}$ &  -$5.33 \times 10^{-4}$  \\ [1.5ex] \cline{1-5} 
                c$_{1L}$ & -5.04 &    -2.10 &   -0.90  &  -0.39 \\ [1.5ex] \cline{1-5} 
                c$_{2L}$ & 0.02&  0.01 &  $9.24 \times 10^{-3}$  & $5.89 \times 10^{-3}$\\ [1.5ex] \cline{1-5} 
                c$_{3L}$ & -6.98  &  -2.66 &  -1.09 & -0.48 \\ [1.5ex] \cline{1-5} 
                c$_{4L}$& 0.34  &  0.17 &  0.09 &  0.05  \\ [1.5ex] \cline{1-5}
                c$_{5L}$ & 0.02 &   $9.67 \times 10^{-3}$ &   $5.89 \times 10^{-3}$ &  $3.60 \times 10^{-3}$  \\ [1.5ex] \cline{1-5} 
			\end{tabular}
\begin{tabular}
{|c|c|c|c|}
				\hline  
                \multicolumn{4}{|c|}{With PNM + $\rho_{c}$(1.8)$\leq$3.5$\rho_0$}\vline
				\\ [1.5ex] \cline{1-4}
				\multicolumn{4}{|c|}{Mass, M (M$_\odot$)}\vline
				\\ [1.5ex] \cline{1-4} 
			      1.2  &  1.4  &  1.6  & 1.8  \\ [1.5ex] \cline{1-4} 
                1109.04 & 451.63  & 192.17 & 81.75 \\ [1.5ex] \cline{1-4} 
                3.20 &  1.57  & 0.80 & 0.42 \\ [1.5ex] \cline{1-4} 
               0.46  &  0.25 & 0.15  & 0.09  \\ [1.5ex] \cline{1-4} 
                9.79 & 3.30      & 1.15 & 0.40    \\ [1.5ex] \cline{1-4} 
                2.03 &  0.93 &  0.43 & 0.20 \\ [1.5ex] \cline{1-4} 
                 -$3.69 \times 10^{-3}$ &  -$1.30 \times 10^{-3}$ &  -$4.33 \times 10^{-4}$ & -$1.15 \times 10^{-4}$ \\ [1.5ex] \cline{1-4} 
                 -$1.17 \times 10^{-3}$ & -$5.40 \times 10^{-4}$ &  -$2.55 \times 10^{-4}$ & -$1.19 \times 10^{-4}$   \\ [1.5ex] \cline{1-4} 
                 0.03  &   0.01 & $6.94 \times 10^{-3}$  &  $3.27 \times 10^{-3}$\\ [1.5ex] \cline{1-4} 
                 -$5.25 \times 10^{-3}$ &   -$1.86 \times 10^{-3}$ &  -$6.43 \times 10^{-4}$ & -$2.09 \times 10^{-4}$\\ [1.5ex] \cline{1-4} 
                 -$7.5 \times 10^{-5}$  & -$4.3 \times 10^{-5}$ &  -$2.6 \times 10^{-5}$ & -$1.7 \times 10^{-5}$ \\ [1.5ex] \cline{1-4} 
                $3.37 \times 10^{-3}$ &  $1.90 \times 10^{-3}$ &   $1.09 \times 10^{-3}$ &  $6.15 \times 10^{-4}$ \\ [1.5ex] \cline{1-4} 
                 -$7.71 \times 10^{-4}$ &  -$3.05 \times 10^{-4}$ & -$1.07 \times 10^{-4}$ & -$1.9 \times 10^{-5}$\\ [1.5ex] \cline{1-4} 
                0.05  &  0.01  &  $5.59 \times 10^{-3}$ & $2.77 \times 10^{-3}$ \\ [1.5ex] \cline{1-4} 
                0.01 &  $4.06 \times 10^{-3}$ &  $2.05 \times 10^{-3}$ &  $1.10 \times 10^{-3}$  \\ [1.5ex] \cline{1-4}
                -$3.18 \times 10^{-3}$ &  -$1.51 \times 10^{-3}$  &   -$7.91 \times 10^{-4}$ &  -$4.63 \times 10^{-4}$  \\ [1.5ex] \cline{1-4} 
                -4.39 &    -1.83 &   -0.80  &  -0.36 \\ [1.5ex] \cline{1-4} 
                 0.02&  0.01 &  $7.71 \times 10^{-3}$  & $4.94 \times 10^{-3}$\\ [1.5ex] \cline{1-4} 
                -7.48  &  -2.98 &  -1.29 & -0.60 \\ [1.5ex] \cline{1-4} 
                0.30  &  0.15 &  0.08 &  0.05  \\ [1.5ex] \cline{1-4}
                0.01 &   $7.95 \times 10^{-3}$ &   $4.88 \times 10^{-3}$ &  $3.02 \times 10^{-3}$  \\ [1.5ex] \cline{1-4} 
\end{tabular}
	\end{table}



\end{document}